\documentclass[sigconf,authorversion]{acmart}
\AtBeginDocument{%
  \providecommand\BibTeX{{%
    \normalfont B\kern-0.5em{\scshape i\kern-0.25em b}\kern-0.8em\TeX}}}

\setcopyright{acmcopyright}
\copyrightyear{2020}
\acmYear{2020}
\acmDOI{}
\acmConference[ASE 2020]{ASE 2020: 35th IEEE/ACM International Conference on Automated Software Engineering}{September 21--25, 2020}{Melbourne, Australia}
\usepackage{xspace}

\usepackage{booktabs}
\usepackage{multirow}
\usepackage{threeparttable}
\usepackage{xcolor}
\usepackage{array,makecell}
\usepackage{diagbox}
\usepackage{tabularx}

\usepackage{adjustbox}
\usepackage{lipsum}
\usepackage[autolanguage]{numprint}
\usepackage[inline]{enumitem}

\def\smartbugs{SmartBugs\xspace}
\def\nbTools{10\xspace}%
\def\nbAnalisis{{428,337}\xspace}%
\def\nbCurated{{143}\xspace}%
\def\nbContracts{{47,518}\xspace}%
\def\ExecDuration{564 days and 3 hours\xspace}
\newcommand\sbset[1]{$\textsc{sb}^{\textsc{#1}}$\xspace}
\def\sbcurated{\sbset{curated}}
\def\sbwild{\sbset{wild}}

\begin{document}

\title{SmartBugs: A Framework to Analyze Solidity Smart Contracts}


\author{Jo\~ao F. Ferreira}
\affiliation{%
    \institution{INESC-ID and IST, University of Lisbon, Portugal}
    %\country{Portugal}
}
\email{joao@joaoff.com}

\author{Pedro Cruz}
\affiliation{%
    \institution{INESC-ID and IST, University of Lisbon, Portugal}
    %\institution{IST, University of Lisbon, Portugal}
    %\country{Portugal}
}
\email{pedrocrvz@gmail.com}

\author{Thomas Durieux}
\affiliation{%
    %\institution{INESC-ID and IST, University of Lisbon, Portugal}
    \institution{KTH Royal Institute of Technology, Sweden}
    %\country{Portugal}
}
\email{thomas@durieux.me}

\author{Rui Abreu}
\affiliation{%
    \institution{INESC-ID and IST, University of Lisbon, Portugal}
    %\country{Portugal}
}
\email{rui@computer.org}

\renewcommand{\shortauthors}{J. F. Ferreira, P. Cruz, T. Durieux, and R. Abreu}

\begin{abstract}
Over the last few years, there has been substantial research on automated analysis, testing, and debugging of Ethereum smart contracts. However, it is not trivial to compare and reproduce that research.
To address this, we present \smartbugs, an extensible and easy-to-use execution framework that simplifies the execution of analysis tools on smart contracts written in Solidity, the primary language used in Ethereum.
\smartbugs\ is currently distributed with support for \nbTools\ tools and two datasets of Solidity contracts. The first dataset can be used to evaluate the precision of analysis tools, as it
contains \nbCurated\ annotated vulnerable contracts with 208 tagged vulnerabilities. The second dataset contains \nbContracts\ unique contracts collected through Etherscan. %extracted from the Ethereum network.
%Users can also create new custom named datasets of smart contracts.
%
We discuss how \smartbugs\ supported the largest experimental setup to date both in the number of tools and in execution time. Moreover, we show how it enables easy integration and comparison of analysis tools by presenting a new extension to the tool SmartCheck that improves substantially the detection of vulnerabilities related to the DASP10 categories \textit{Bad Randomness}, \textit{Time Manipulation}, and \textit{Access Control} (identified vulnerabilities increased from 11\% to 24\%).
\end{abstract}



%%\begin{CCSXML}
%%<ccs2012>
%%<concept>
%%<concept_id>10011007.10011074.10011099.10011102</concept_id>
%%<concept_desc>Software and its engineering~Software defect analysis</concept_desc>
%%<concept_significance>500</concept_significance>
%%</concept>
%%<concept>
%%<concept_id>10011007.10011074.10011099.10011102.10011103</concept_id>
%%<concept_desc>Software and its engineering~Software testing and debugging</concept_desc>
%%<concept_significance>500</concept_significance>
%%</concept>
%%</ccs2012>
%%\end{CCSXML}

%%\ccsdesc[500]{Software and its engineering~Software defect analysis}
%%\ccsdesc[500]{Software and its engineering~Software testing and debugging}



\maketitle

%%%%%%%%%%%%%%%%%%%%%%%%%%%%%%%%%%%%%%%%%%%%%%%%%%%%%%%%%%%%%%%%%
%%%%%%%%%%%%%%%%%%%%%%%%%%%%%%%%%%%%%%%%%%%%%%%%%%%%%%%%%%%%%%%%%
\section{Introduction}
%Blockchain technology has been receiving considerable attention from industry and academia, for it promises to disrupt the digital online world by enabling a democratic, open, and scalable digital economy based on decentralized distributed consensus without the intervention of third-party trusted authorities.
%Among the currently available blockchain-based platforms,
Ethereum is one of the most popular blockchain-based platforms, mainly because it enables developers to write distributed applications (Dapps) based on smart contracts\,---\,programs that are executed across a decentralised network of  nodes. The main language used to develop Ethereum smart contracts
is Solidity\footnote{Interested readers on Solidity, refer to 
\url{https://solidity.readthedocs.io}.},
a high-level language that follows a 
JavaScript-like, object-oriented paradigm. 
Contracts written in Solidity are compiled to bytecode that can be executed on the Ethereum Virtual Machine (EVM).

Smart contracts are at the core of Ethereum's value. 
However, as noted by some researchers~\cite{bhargavan2016formal,luu2016making}, writing secure smart contracts is far from trivial.
In a preliminary study performed on nearly one million Ethereum smart contracts, using one 
analysis framework for verifying correctness, 
\emph{34,200} of them were flagged as vulnerable~\cite{nikolic2018finding}. 
%Also, Luu \textit{et~al.}~\cite{luu2016making} proposed the symbolic execution tool Oyente and
%showed that of \emph{19,366} Ethereum smart contracts analyzed, \emph{8,833} (around 46\%) were flagged as vulnerable. 
Famous attacks, such as TheDAO exploit\footnote{Analysis of the DAO exploit (Phil Daian): \url{https://bit.ly/2XOqVmy}}
%~\cite{analysisthedao} 
and the Parity wallet bug\footnote{The \$280M Ethereum’s Parity bug (Matt Suiche): \url{https://bit.ly/3guX8Yx}}
%~\cite{paritywallet} 
illustrate this problem and have led to huge financial 
losses.

% problem
There has been some effort from the research community to develop automated
analysis tools that locate and eliminate vulnerabilities in smart contracts~\cite{luu2016making,tikhomirov2018smartcheck,tsankov2018securify,grishchenko2018semantic}. 
However, it is not easy to compare and reproduce that research:
even though several of the tools are publicly available, the datasets used are not.
If a developer of a new tool wants to compare the new tool with
existing work, the current approach is to contact the authors of alternative
tools and hope that they give access to their datasets (as done in, e.g.,~\cite{perez2019smart}).

The aim of this paper is to present \smartbugs, 
an extensible and easy-to-use execution framework that simplifies the execution of analysis tools on Solidity smart contracts and facilitates reproducibility.
We describe the architecture of the framework, the tools and datasets provided, and the methodologies used for adding new tools and for filtering datasets (\S\ref{sec:smartbugs}). We illustrate two typical use cases where \smartbugs\ can be used (\S\ref{sec:use-cases}). First, we discuss how it supported
the largest experimental setup to date both in the number of tools and in execution time~\cite{durieux2019empirical}. Second, we show how it can be used to compare tools by adding a new extension of SmartCheck~\cite{tikhomirov2018smartcheck}
that improves substantially the detection of vulnerabilities related to the DASP10 categories \textit{Bad Randomness}, \textit{Time Manipulation}, and \textit{Access Control} (identified vulnerabilities increased from 11\% to 24\%).

\smartbugs\ is open-source and is publicly available online at:
\begin{center}
\bf
    \url{https://smartbugs.github.io}
\end{center}

%%%%%%%%%%%%%%%%%%%%%%%%%%%%%%%%%%%%%%%%%%%%%%%%%%%%%%%%%%%%%%%%%
%%%%%%%%%%%%%%%%%%%%%%%%%%%%%%%%%%%%%%%%%%%%%%%%%%%%%%%%%%%%%%%%%
\section{SmartBugs}\label{sec:smartbugs}
\begin{figure}
   \includegraphics[width=0.9\columnwidth]{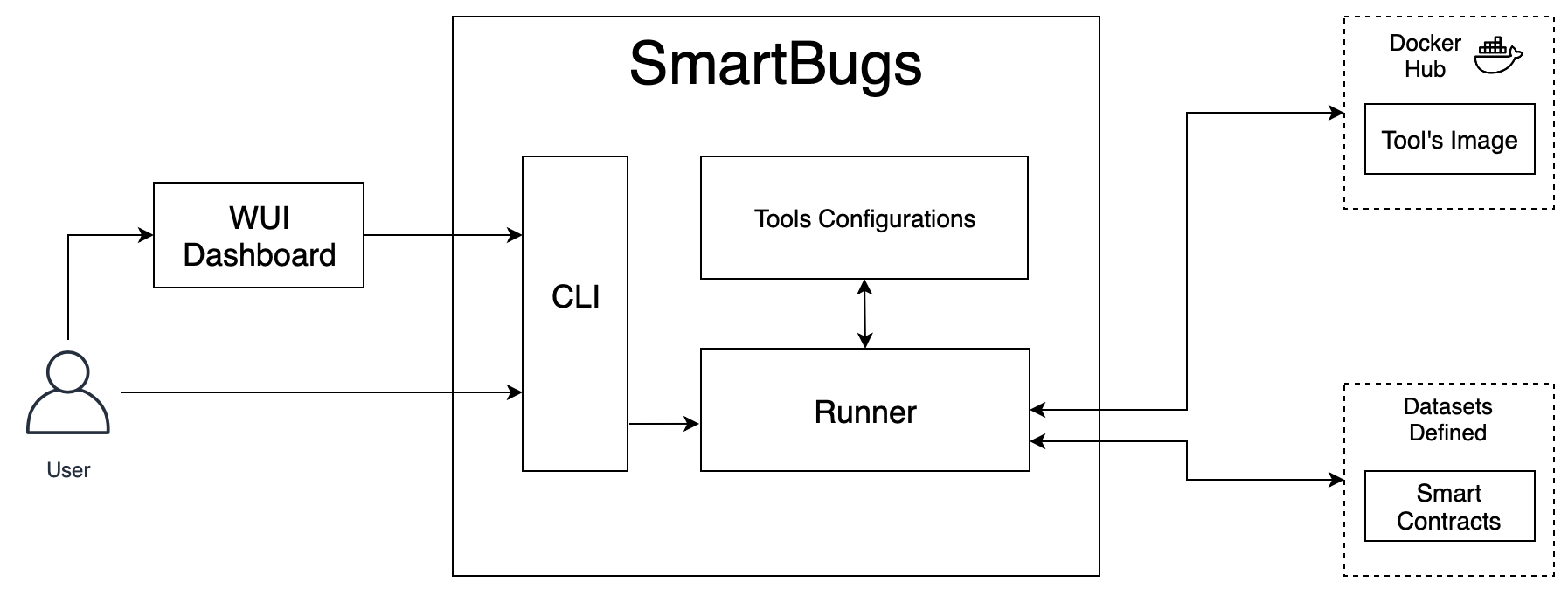}
   \caption{SmartBugs Architecture\label{pic:smartbugs-arch}}
\end{figure}

This section describes \smartbugs, focusing on system requirements, available tools and datasets, methodologies for adding tools and filtering datasets, and the available interfaces. \smartbugs is composed of five main parts: the command-line interface, the tool configurations, the Docker images of the tools, the datasets of smart contracts, and the \smartbugs\ Runner, which brings all the parts together to execute the analysis tools.
%\begin{enumerate*}
%\item The first consists of the command-line interface. % to use \smartbugs. %(see \autoref{sec:smartbugs:interface}).
%\item The second part contains the tool configurations. %Each tool plugin contains the configuration of the tools. 
%%The configuration contains the name of the Docker image, the name of the tool, the command line to run the tool, the description of the tool, and the location of the output of results.
%\item The Docker images of the tools stored on Docker Hub. We use pre-existing Docker images when available; otherwise, we create our own image (all Docker images are made publicly available on Docker Hub).
%\item The datasets of smart contracts. At the time of writing, \smartbugs\ is distributed with two datasets (\nbContractsAll\ contracts in total). % (see \autoref{sec:smartbugs:datasets}).
%\item The \smartbugs' Runner, which brings all the parts of \smartbugs\ together to execute the analysis tools.
%\end{enumerate*}
We also provide a web-based user interface that interacts with \smartbugs. Figure~\ref{pic:smartbugs-arch} shows how the different \smartbugs\ components are put together.

%%%%%%%%%%%%%%%%%%%%%%%%%%%%%%%%%%%%%%%%%%%%%%%%%%%%%%%%%%%%
\subsection{System Requirements}

SmartBugs requires Docker and Python3 with the modules
\textit{PyYAML}, \textit{solidity\_parser}, and \textit{docker}.
Since Solidity versions are not always backwards-compatible, the 
%tools currently available 
analysis tools
might have problems processing some contracts depending on the solidity compiler used. For example, Solidity v0.5.0 introduced breaking changes\footnote{Solidity v0.5.0 introduced breaking changes: \url{https://bit.ly/2W0bY0x}}
%\url{https://solidity.readthedocs.io/en/v0.5.0/050-breaking-changes.html}} 
and this creates compatibility issues with some versions of the Mythril tool.
%Indeed, Mythril and Securify have compatibility issues with specific Solidity versions since they analyse directly the EVM bytecode.
%However, tools such as SmartCheck that analyse the source code are less impacted by the changes between versions. 

To mitigate this problem, \smartbugs\ provides the possibility of having two different versions of the same tool by adding a property in the configuration file. The configuration file supports a default tool version to compile and analyse contracts above Solidity v0.5.0 (or all contracts if no other tool version is provided). It is also possible to specify a different tool version to compile and analyse contracts below Solidity v0.5.0. 
This is illustrated in Section~\ref{add-tools}.
%For example, tools such as Mythril and Securify, have two distinct versions configured: one for contracts with Solidity below v0.5.0 and other for contracts written with Solidity above v0.5.0.

%{\color{red}
%The design decision that we took was to put the users in control, allowing them to adapt the dataset to their own needs. 
%}

%%%%%%%%%%%%%%%%%%%%%%%%%%%%%%%%%%%%%%%%%%%%%%%%%%%%%%%%%%%%
\subsection{Available Tools and Datasets}
At the time of writing, \smartbugs\ comes with \nbTools{} tools ready to be used: HoneyBadger, Maian, Manticore, Mythril, Osiris, Oyente, Securify, Slither, SmartCheck, Solhint. %{\color{red} TODO: refs}
%
%The framework and \sbcurated are available in GitHub as repositories of \smartbugs~\footnote{SmartBugs (including \sbcurated): \url{https://github.com/smartbugs/smartbugs}}\footnote{SmartBugs WUI: \url{https://github.com/smartbugs/smartbugs-dashboard}}.
%
It is also distributed with
two datasets of Solidity contracts. The first dataset is named \sbcurated\ and contains \nbCurated\ annotated vulnerable contracts with 208 tagged vulnerabilities, divided into 10 categories. This dataset can be used to evaluate the precision of analysis tools. The second dataset is named \sbwild\ and it contains \nbContracts\ unique\footnote{We consider two contracts to be duplicates when their MD5 checksum is the same after removing all the spaces and tabulations.} contracts collected through Etherscan. 
All contracts and tools are publicly available. The collection methodology for \sbcurated\ is explained in this section. For details about \sbwild, we refer the reader to~\cite{durieux2019empirical}.

Our objective in constructing \sbcurated\ is to provide a reliable dataset with a collection of vulnerabilities designed to be reproducible, that follows a known taxonomy and that can serve as a reference dataset to the research community. The dataset follows the taxonomy of DASP 10.\footnote{DASP 10 taxonomy: \url{https://dasp.co}} Since the category \textit{Unknown Unknowns} represents future and undiscovered vulnerabilities, 
%This category is not useful in the context of mapping existent vulnerabilities, so 
we opted to map vulnerabilities that did not fit any other of the nine categories into this category (e.g. vulnerabilities such as uninitialized data and the possibility of locking down Ether). For simplicity, we use the nomenclature \textit{Other} instead of \textit{Unknown Unknowns}. %will be referred as \textit{Other} to avoid confusion and simplify the description.

\sbcurated was created by collecting smart contracts from three different sources: GitHub repositories, Blog posts that analyse contracts and the Ethereum network. Most of contracts were collected from GitHub repositories and the Ethereum network. We ensure the traceability of each contract by providing the URL from which they were taken and its author, where possible.
%
%The dataset contains \nbCurated\ contracts and 208 manually tagged vulnerabilities, divided into 10 categories of vulnerabilities.
Table~\ref{tab:selected-categories} shows how the \nbCurated\ contracts are categorized. Each row contains a category of vulnerability. For each category, we provide the number of contracts available within that category and the total number of vulnerabilities and number of lines of code of the contracts of that category.

\begin{table}
\small
    \caption{Categories of vulnerabilities available in the dataset~\sbcurated. LoC computed using cloc 1.82.}
    \label{tab:selected-categories}
    \begin{center}
    \begin{tabular}{|p{0.21\textwidth}|r|r|r|} \hline
            \textbf{Category} & \textbf{Contracts} & \textbf{Vulns} & \textbf{LoC}  \\
         \hline
Access Control & 17 & 19 & \numprint{899}  \\
 
Arithmetic & 14 & 24 &\numprint{304}  \\

Bad Randomness & 8 & 30 &\numprint{1079} \\

Denial of service & 6 & 7 & 177  \\
 
Front running & 4 & 7 & 137 \\
 
Reentrancy & 31 & 32 & \numprint{2164} \\

Short addresses & 1 & 1 & 18 \\

Time manipulation & 5 & 7 & 100 \\

Unchecked low level calls &  53 & 78  & 4055 \\

Other  & 3  & 3 & 115 \\
         \hline
         \textbf{Total} & 143 & 208 & \numprint{9048} \\
    \hline
    \end{tabular}
    \end{center}
\end{table}

%\sbcurated design was driven by the following objectives: \textit{Reproducibility}, \textit{Real-world relevance}, \textit{Diversity} and \textit{Custom filtering}.

%%%%%%%%%%%%%%%%%%%%%%%%%%%%%%%%%%%%%%%%%%%%%%%%%%%%%%%%%%%%
\subsection{Methodology for Adding Tools}\label{add-tools}
Addition of tools in SmartBugs is designed to be simple and practical, allowing the user to control the execution of the tools according to their needs. Currently, all the tools in \smartbugs\ use Docker images pulled from Docker Hub. %Some images were already publicly available, whilst others were created by us. %to be added to \smartbugs.
We use pre-existing Docker images when available; otherwise, we create our own image (all Docker images are made publicly available on Docker Hub).
The choice to use Docker images was made to ease the addition of tools, allow the execution to be reproducible and use the same execution environment for all tools, allowing the user to execute \smartbugs\ in any environment where Python3 and Docker are installed.

Each tool plugin contains the configuration of the tool. The configuration contains the name of the Docker image, the name of the tool, the command to run the tool, and, optionally, the description of the tool and the location of the output of results. Once a Docker image providing the tool is available, adding the tool to \smartbugs\ consists of adding a new configuration file (an \textit{.YAML} file) such as the following:   
%with the following structure:
%{\small
%\begin{verbatim}
%    docker_image:
%        default: qspprotocol/mythril-usolc
%        solc<5: qspprotocol/mythril-0.4.25
%    cmd: -x
%    info: Mythril analyses EVM bytecode using symbolic analysis,
%    taint analysis and control flow checking to detect a variety
%    of security vulnerabilities.
%\end{verbatim}
%\label{verb:struct}
%}
{\small
\begin{verbatim}
    docker_image:
        default: qspprotocol/securify-usolc
        solc<5: qspprotocol/securify-0.4.25
    cmd: --livestatusfile /results/output.json -fs
    
    output_in_files:
      folder: /results/output.json
\end{verbatim}
\label{verb:struct}
}
%info: Securify uses formal verification, also relying
%on static analysis checks. Securify’s analysis consists
%    info: Securify’s analysis consists of two steps. First, it
%    symbolically analyses the contract’s dependency graph to
%    extract precise semantic information from the code. Then, 
%    it checks compliance and violation patterns that capture
%    sufficient conditions for proving if a property holds.
By default, \smartbugs extracts the results from the output printed by each tool. If instead a tool stores the result of the analysis in a file in the Docker image, the path of that file should be defined in the configuration file using the optional configuration parameter \verb@output_in_files@, as shown above.

Finally, when adding a tool to \smartbugs, a \verb@parse@ method can be implemented so that the output with the vulnerabilities detected by the tool is normalized.\footnote{For example, the parser for SmartCheck is defined here: \url{https://bit.ly/3exxeRW}.}

%%%%%%%%%%%%%%%%%%%%%%%%%%%%%%%%%%%%%%%%%%%%%%%%%%%%%%%%%%%%
\subsection{Methodology for Filtering Datasets}\label{filter-bugs}
\smartbugs supports the definition of \textit{named datasets}, which represent subsets of contracts that share a common property. For example, a named dataset already provided by default is \textit{reentrancy}: it corresponds to contracts that are identified as being vulnerable to reentrancy attacks. Named datasets can be specified in a configuration file (\textit{config/dataset/dataset.yaml}). %The default configuration file provides the named sets shown in Table~\ref{tab:selected-categories}.
To add a custom named dataset, the user simply has to alter the configuration file by adding a name and the correspondent list of paths. The path can be a directory, a file, or a list of both. For example:

{\small
\begin{verbatim}
  reentrancy: dataset/reentrancy
  arithmetic:
    - dataset/arithmetic
    - dataset/reentrancy/reentrance.sol
\end{verbatim}
\label{verb:structdataset}
}
%The dataset configuration file defines the name of the datasets and the path to the correspondent directories or files.

%%%%%%%%%%%%%%%%%%%%%%%%%%%%%%%%%%%%%%%%%%%%%%%%%%%%%%%%%%%%
\subsection{Command-Line Interface}
\smartbugs\ provides a command line interface that allows users to run different analysis tools on the available datasets of contracts. The user can also get information about the tools, if provided, skip an execution that already has results, specify the number of processes to use during the analysis (by default 1) and list the named datasets and tools available.
\smartbugs\ command-line interface can be invoked as:

{\small
\begin{verbatim}
  smartBugs.py  [-h, --help]
                (--file FILES | --dataset DATASETS) 
                --tool TOOLS --info TOOLS                     
                --skip-existing --processes PROCESSES
                --list {tools, datasets}                
\end{verbatim}
}

\noindent \textbf{Usage Example}
%The following example shows how we can use the CLI of \smartbugs\
To run the tools Oyente and Mythril against the contracts in the named dataset \textit{reentrancy}, we can execute:

{\small
\begin{verbatim}
  smartBugs.py --tool oyente mythril --dataset reentrancy
\end{verbatim}
}
\noindent This command creates an output folder with the results of the analysis for each tool executed. By inspecting the output files, we can determine very quickly which contracts are identified as having vulnerabilities.
%that both tools found problems in exactly the same contracts, further increasing confidence that the analyses performed by these tools are correct. 
Since all the tools added to \smartbugs\ come with a parser mechanism to normalize the output, a \texttt{json} file, with all vulnerabilities detected by the tool is created. A file containing the raw output of the tool executed is also generated in the same folder. Also, the \smartbugs logs are stored in a folder called \texttt{logs} composed of files named with the date and hour of the execution.

\subsection{WUI Dashboard}
We also provide a Web-based UI (WUI) that interacts with \smartbugs.\footnote{\smartbugs\ Dashboard: \url{https://github.com/smartbugs/smartbugs-dashboard}} This dashboard provides the user easy access to the list of tools, named datasets available and the vulnerabilities detected by each tool available mapped to a category of DASP 10. 
Figure~\ref{fig:wui-smartbugs} shows a screenshot of the dashboard. It offers three options to analyse smart contracts:
\begin{figure}
   \includegraphics[width=0.9\columnwidth]{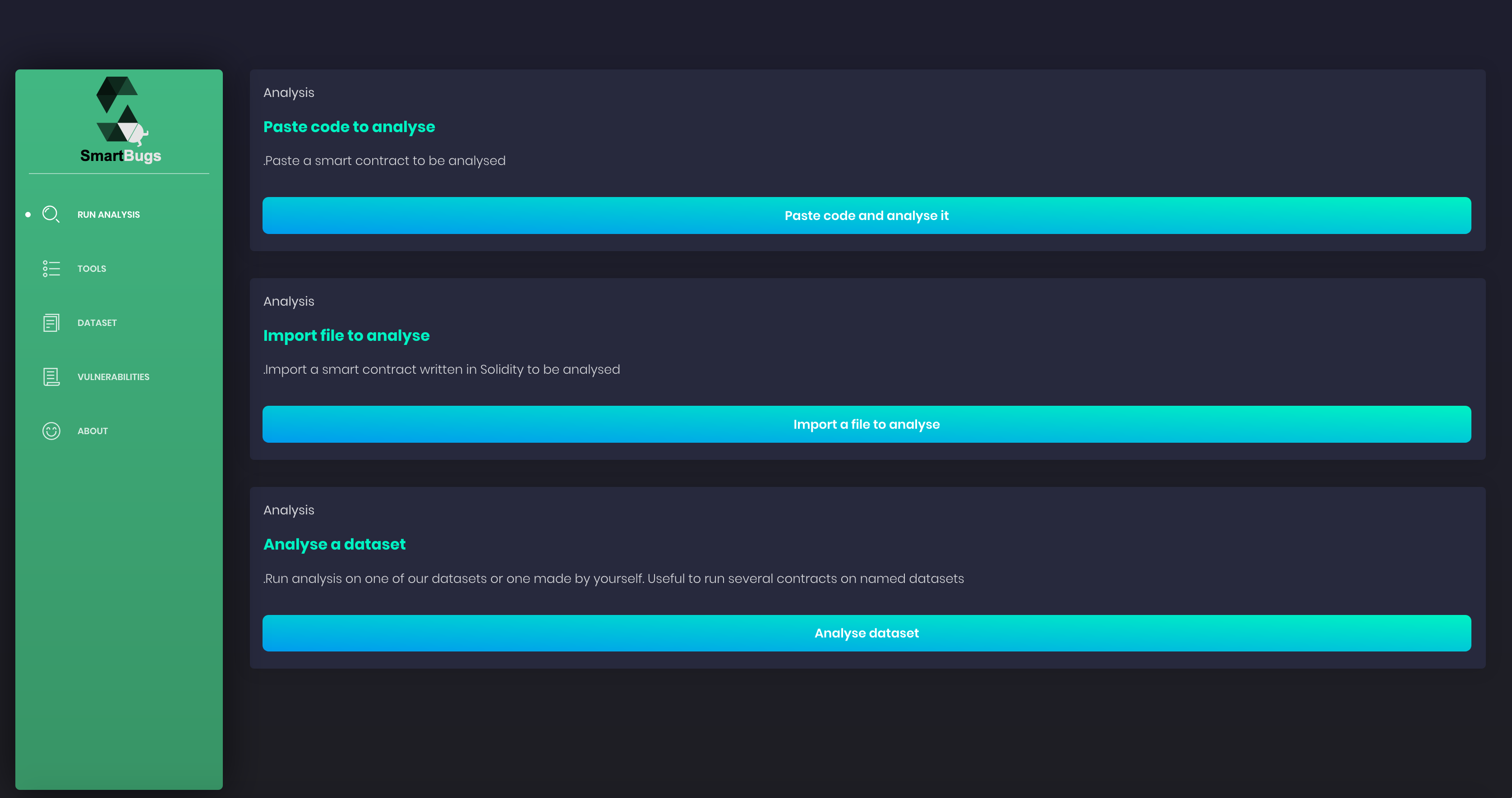}
   \caption{SmartBugs Web Dashboard\label{fig:wui-smartbugs}}
\end{figure}
%SmartBugs WUI is not dynamically updated, meaning that if new tools or datasets are added to SmartBugs, they will not be automatically available on the WUI (unlike CLI). The reason for this is that there is a need to compute the outputs and thus a parser mechanism must be defined in order to present results to the user. Instead, to add a tool or dataset, first the user should follow the procedure explained in \ref{add-tools} and \ref{filter-bugs} and then manually update the WUI by adding the name of the tool or the dataset. It should also be defined a way to parse the output. The dashboard currently supports 10 tools, includes \sbcurated and provides a detailed mapping of all vulnerabilities identified by the 10 tools to a DASP 10 category.
%
\begin{enumerate*}
    \item The user can paste or write a smart contract directly in the browser;
    \item The user can import a smart contract by uploading a file;
    \item The user can run the available tools on pre-defined datasets (from \sbcurated).
\end{enumerate*}
After execution, the dashboard shows a graph with the number of security issues found by each tool, and for each tool it presents the issues found.

%\noindent \textbf{Usage Example} The following figures show how we can use the WUI of SmartBugs to run the tools Oyente, Slither and SmartCheck against the contract inputted and how the results are shown:
%\begin{center}
%   \includegraphics[scale=0.33]{Images/smartbugs-ui4.png}
%   \captionof{figure}{Input panel for smart contracts and tool's choice (Oyente, Slither and SmartCheck are selected)}
%\end{center}

%\begin{center}
%   \includegraphics[scale=0.5]{Images/smartbugs-ui1.png}
%   \captionof{figure}{Number of issues found by each tool}
%\end{center}

%\begin{center}
%   \includegraphics[scale=0.3]{Images/smartbugs-ui2.png}
%   \captionof{figure}{Issues found by Oyente}
%\end{center}

%For each tool executed, a similar output to Figure 3.5 is presented.

%%%%%%%%%%%%%%%%%%%%%%%%%%%%%%%%%%%%%%%%%%%%%%%%%%%%%%%%%%%%%%%%%
%%%%%%%%%%%%%%%%%%%%%%%%%%%%%%%%%%%%%%%%%%%%%%%%%%%%%%%%%%%%%%%%%
\section{Use Cases}\label{sec:use-cases}
The primary envisaged users of \smartbugs\ are researchers who are interested in automated analysis and debugging of Solidity smart contracts. In this section, we present two typical use cases. First, we summarize an empirical evaluation that was supported by \smartbugs~\cite{durieux2019empirical}. We then show how \smartbugs\ can support tool developers by discussing how a new extension of SmartCheck~\cite{tikhomirov2018smartcheck} can be easily compared with the original tool.

\subsection{Supporting Empirical Evaluations}\label{section:use-case-empirical}
\smartbugs\ can support researchers who are interested in doing large empirical evaluations. The command-line interface and the options \verb@--skip-existing@ and \verb@--processes@ are particularly helpful. We have recently used \smartbugs\ to 
obtain an overview of the current status of automated analysis tools for Solidity smart contracts 
and to
support
the largest experimental setup to date both in the number of tools and in execution time~\cite{durieux2019empirical}.
We evaluated \nbTools state-of-the-art automated analysis tools on \sbwild\ and on a subset of \sbcurated\ that contained 69 contracts (since then, the number of contracts in \sbcurated\ has increased). In total, we ran \nbAnalisis\ analyses that took approximately \ExecDuration. %, being the largest experimental setup to date both in the number of tools and in execution time.
We found that only 42\% of the vulnerabilities from the annotated dataset are detected by all the tools, with the tool \textit{Mythril} having the higher accuracy (27\%).
When considering the largest dataset, \sbwild, we observed that 97\% of contracts are tagged as vulnerable, thus
suggesting a considerable number of false positives. %Indeed, only a small number of vulnerabilities (and of only two categories) were detected simultaneously by four or more tools.
%{\color{red} TODO: punch line about use of smartbugs}

The use of \smartbugs\ made the task easier and was crucial to ensure that the work can be completely reproduced.

%%%%%%%%%%%%%%%%%%%%%%%%%%%%%%%%%%%%%%
%\subsection{SmartCheck Extension}
\subsection{Supporting Developers of Analysis Tools}
The empirical evaluation described above showed that there is room for improvement for automated analysis tools to detect more vulnerabilities. For example, \textit{Bad Randomness} was one of the categories that all of the tools failed to detect. In this section, we describe a simple extension of the tool SmartCheck~\cite{tikhomirov2018smartcheck} that enables the detection of vulnerabilities related to \textit{Bad Randomness} and improves detection of \textit{Time Manipulation} and \textit{Access Control} vulnerabilities.\footnote{Descriptions of these vulnerabilities can be found in DASP's website: \url{https://dasp.co}} We refer to our extension as SmartCheck Extended.

SmartCheck runs lexical and syntactical analysis on Solidity source code. It uses a custom Solidity grammar to generate an XML parse tree as an intermediate representation (IR). SmartCheck detects vulnerability patterns by using XPath patterns on the IR. Our approach to improve SmartCheck vulnerabilities detection was to add new rules, in the form of XPath patterns. %One of the problems of the SmartCheck approach, using XPath patterns, is that more complex rules can not be precisely described using XPath. So, when more complex cases are needed to be put in form of XPath patterns they can easily lead to false positives.

We added three new rules to SmartCheck. 
%to improve an already established rule. Our first approach, based on the analysis presented in Chapter \ref{empirical-analysis}, was to add a rule to detect Bad Randomness issues called \textit{SOLIDITY\_BAD\_RANDOMNESS}. To do that, 
The first rule is named \textit{SOLIDITY\_BAD\_RANDOMNESS} and aims at detecting issues related to the category \textit{Bad Randomness}. For this, we created an XPath pattern to detect the use of environment variables such as \textit{block.number}, \textit{block.coinbase}, \textit{block.difficulty}, \textit{block.gaslimit}, \textit{blockhash}, and \textit{block.blockhash}. For the second rule, we followed a similar approach to update the rule \textit{SOLIDITY\_EXACT\_TIME}, already included in SmartCheck. We modified the pattern to look for expressions that contain \textit{block.timestamp} or \textit{now}, extending the previously defined rule for cases more general than comparisons. %not only when used in comparisons as previously defined. 
These rules are straightforward lexical analyses whose goal is to simply detect the use of the referred environment variables and to flag their use, acting as a warning.

Regarding \textit{Access Control}, SmartCheck's default rule is restricted to \textit{tx.origin} issues. To improve this, we added a pattern to search for `suicides' (uses of \textit{selfdestruct}) and ownership transfers where the function misses proper protection. We constructed two rule patterns inside a single rule named \textit{SOLIDITY\_UNPROTECTED}. To detect unprotected issues we created a pattern to look for all functions defined, excluding constructors, that do not have standard \textit{modifiers} defined, as \textit{onlyOwner}, or \textit{require} statements protecting a value assignment to a variable defined as \textit{owner} or \textit{selfdestruct} calls.

The source code of SmartCheck Extended is available on GitHub\footnote{SmartCheck Extended: \url{https://github.com/pedrocrvz/smartcheck}} as a fork of the original SmartCheck. It is also included in \smartbugs\ and ready to executed.

\subsubsection{Results}
We used \smartbugs\ to compare our extension with the original tool. 
%\sbcurated with SmartCheck Extended, using the same version used in the study presented in Chapter~\ref{chap:analysis}, to set a fair ground of comparison. 
Table \ref{tab:smartbugs:flagged-categories-extension} compares the results obtained for SmartCheck in the empirical evaluation described in Subsection~\ref{section:use-case-empirical} with the results obtained from executing our extension on the same dataset of contracts. 
The results shown in the table only consider the 69 contracts used in the empirical study mentioned above~\cite{durieux2019empirical}, so that we can perform a fair comparison.
We can observe that SmartCheck Extended is capable of detecting a total of 15 more issues, %with regard of the annotated vulnerabilities in \sbcurated, 
more than doubling the capability of detection when compared to SmartCheck. With our proposed extension we can detect 24\% of the vulnerabilities annotated in \sbcurated, instead of the previous 11\%.
More details about this extension, including evaluation on its precision, are presented in~\cite{cruz-extended-abstract}.

\begin{table}

    \caption{Vulnerabilities identified per category by SmartCheck and SmartCheck Extended in \sbcurated}
    \label{tab:smartbugs:flagged-categories-extension}
      \begin{adjustbox}{width=0.49\textwidth}
    \centering
\small
    \begin{tabular}{|l|r|r|} 
    \hline
\textbf{Category}                    & \textbf{SmartCheck}   & \textbf{SmartCheck Extended}         \\\hline
Access Control              &  2/19 11\%   &    4/19  21\% \\
Arithmetic                  &    1/22  5\% &    1/22  5\% \\
Bad Randomness              &    0/31  5\% &    10/31 32\% \\
Denial of Service           &    0/7   0\% &     0/7   0\% \\
Front Running               &    0/7   0\% &     0/7  0\% \\
Reentrancy                  &    5/8  62\% &    5/8  62\% \\
Short Addresses             &    0/1  0\% &    0/1  0\% \\
Time Manipulation           &    1/5  20\% &    4/ 5  80\% \\
Unchecked Low~Level~Calls   &   4/12  33\% &    4/12  33\% \\
Other                       &    0/3   0\% &    0/3 0\% \\\hline
\textbf{Total}                       &  13/115 11\% &.    28/115 24\% \\
    \hline
    \end{tabular}
    
    \end{adjustbox}
\end{table}

\section{Conclusion}\label{sec:conclusion}
This paper presents \smartbugs, an extensible and easy-to-use execution framework that simplifies the execution of analysis tools on Solidity smart contracts. One of the main goals of \smartbugs is to facilitate the reproducibility of research in automated reasoning and testing of smart contracts. To demonstrate that integration of new tools and comparison with existing tools is easy, we extended SmartCheck and used \smartbugs\ to show that our extended version improves substantially the detection of vulnerabilities related to Bad Randomness, Time Manipulation, and Access Control.

We believe that \smartbugs\ can be a valuable asset for driving research in automated analysis of smart contracts.
%and we plan to keep improving it. Some of the next steps 
Future work includes i) addition of new analysis tools, ii) expansion of the datasets with more contracts, iii) improved documentation (e.g. contribution guidelines), and iv) new empirical studies supported by \smartbugs.

{
%\footnotesize
\begin{acks}
This work has been co-funded by the European Union's Horizon 2020 research and innovation programme under the QualiChain project, Grant Agreement No 822404 and 
    supported by national funds through FCT, Funda\c{c}\~ao para a Ci\^encia e a Tecnologia, under projects UIDB/50021/2020 and PTDC/CCI-COM/29300/2017.
\end{acks}
}   
\bibliographystyle{ACM-Reference-Format}
\bibliography{references}

\end{document}